\documentstyle[prd,aps,preprint,psfig]{revtex}

\begin{document}


\title{Observational Implications of Axionic Isocurvature Fluctuations}

\author{Toshiyuki Kanazawa$^1$, M. Kawasaki$^2$, Naoshi Sugiyama$^3$
and T. Yanagida$^{1,2}$} 
\address{$^1$Department of Physics, University of Tokyo, Tokyo
113-0033, Japan}
\address{$^2$Research Center for the Early Universe (RESCEU),
University of Tokyo, Tokyo 113-0033, Japan}
\address{$^3$Department of Physics, Kyoto University, Kyoto 606-8502,
Japan}
\date{\today}

\maketitle
\begin{abstract}
The axion is the most attractive candidate to solve the strong
  CP problem in QCD. If it exists, the inflationary universe produces
  axion fluctuations which are mixtures of isocurvature and adiabatic
  fluctuations in general. We investigate how large isocurvature
  fluctuations are allowed or favored in order to explain observations
  of the large scale structure of the present universe.  Generic flat
  universe models with mixed (isocurvature$+$adiabatic) density
  fluctuations are studied.  It is found that the observations are
  consistent with the mixed fluctuation model if the ratio $\alpha$ of
  the power spectrum of isocurvature fluctuations to that of adiabatic
  fluctuations is less than $\sim 0.1$. In particular, the mixed
  fluctuation model with $\alpha \sim 0.05$, total matter density
  $\Omega_{0} =0.4$, and Hubble parameter $H_0=70$km/s/Mpc gives a
  very good fit to the observational data.  Since the height of the
  acoustic peak in the angular power spectrum of the cosmic microwave
  background (CMB) radiation decreases significantly when the
  isocurvature fluctuations are present, the mixed fluctuation model
  can be tested in future satellite experiments. Ratios of the
  amplitude at the peak location to that at the COBE normalization
  scale for various models are given.  Furthermore, we also obtain the
  amplitude of isocurvature fluctuations as a function of axion
  parameters and the Hubble parameter during the inflation. We discuss
  the axion fluctuations in some realistic inflation models and find
  that a significant amount of the isocurvature fluctuations are
  naturally produced.
\end{abstract}
\clearpage

\section{Introduction}

The axion is a Nambu-Goldstone boson associated with spontaneous
breaking of the Peccei-Quinn $U(1)$ symmetry which is introduced in
order to solve the strong CP problem in quantum chromodynamics (QCD).
The breaking scale $F_{a}$ of the Peccei-Quinn symmetry is stringently
constrained by accelerator experiments, astrophysics and cosmology,
and it should lie in the rage $10^{10} - 10^{12}$GeV in standard
cosmology. In particular, the upper limit is set by requiring that the
cosmic density of the axion does not exceed the critical density of
the universe. In other words, the axion should represent the dark
matter of the universe if $F_{a}$ is $\sim 10^{12}$GeV, which makes
the axion very attractive.

In the inflationary universe, quantum fluctuations of the inflaton
field result in adiabatic density fluctuations with a scale-invariant
power spectrum which would account for the large scale structure of
the present universe.  This natural generation of density fluctuations
is one of successes of the inflationary universe.  However, if the
axion exists, another kind of fluctuations is produced in the
inflationary universe.  During inflation the axion has quantum
fluctuations whose root mean square amplitude $\delta a$ is given by
$H/(2\pi)$. ($H$ is the Hubble parameter during inflation.)  The
quantum fluctuations of the axion are stretched by the inflation and
become classical.  When the axion acquires a mass $m_{a}$ at the QCD
scale, these axion fluctuations result in density fluctuations of the
axion.  However, since the axion is massless during inflation, axion
fluctuations with wavelength larger than the horizon size do not
contribute to the total density fluctuations. For this reason, such
density fluctuations are called ``isocurvature''.  Therefore, if
indeed the dark matter consists of axions, it has both adiabatic and
isocurvature fluctuations and may play an important role in the
structure formation of the universe.

In a previous work~\cite{KSY} the large scale structure formation due
to the mixed (adiabatic $+$ isocurvature) fluctuations of the axion
was studied with matter density parameter $\Omega_{0} =1$.  It was
found that the introduction of isocurvature fluctuations significantly
reduces the amplitude of the power spectrum $P(k)$ of the density
fluctuations, keeping $P(k)$ normalized by the COBE data.  Since it is
known that the cold dark matter (CDM) model with pure adiabatic
density fluctuations and $\Omega_{0}=1$ predicts an amplitude of
$P(k)$ that is too large on galaxy and cluster of galaxies scales, the
mixed fluctuation model gives a better fit to the observations,
although the shape $P(k)$ on larger scales does not quite fit the
observations.  Furthermore, it was also pointed out that the height of
the acoustic peak in the angular power spectrum of the cosmic
microwave background (CMB) radiation decreases when the isocurvature
fluctuations are added, which can be tested in the future satellite
experiments. Similar studies were also done in Refs. 2) and
3).\footnote{A similar result was obtained in the context of the hot
  dark matter model.~\cite{Scherrer}}


However, previous works deal with only restricted cosmological models
(i.e. $\Omega_{0} =1$).  It is well known that the shape of the power
spectrum inferred from the galaxy survey~\cite{Peacock} favors a low
matter density universe.  Therefore, in this paper we consider the
structure formation with the mixed fluctuations in a generic flat
universe (i.e.  $\Omega_{0} + \lambda_{0} = 1$, where $\lambda_{0}$ is
the density parameter for the cosmological constant) and investigate
how large isocurvature fluctuations are allowed or favored in order to
explain the observations.  The constraint on the amplitude of the
isocurvature fluctuations is reinterpreted as a constraint on the
Hubble parameter $H$ and the Peccei-Quinn scale $F_{a}$.  We also
study the cosmological evolution of the axion fluctuations in some
inflation models. Among many inflation models, we adopt a hybrid
inflation model and a new inflation model. In the hybrid inflation
model, which is most natural in supergravity, we find that a
significant amount of isocurvature fluctuations of the axion are
naturally produced if $F_a \simeq 10^{15-16}$~GeV. We note that
late-time entropy production may easily raise the $F_a$ up to $\sim
10^{16}$GeV. In the new inflation model, we also find that
isocurvature fluctuations are produced with a canonically favored
value of $F_a \sim 10^{12}$GeV without late-time entropy production.

In \S\ref{sec:obs} we investigate the structure formation with mixed
fluctuations and compare the theoretical predictions with the present
observations. The generation of isocurvature and adiabatic
fluctuations of the axion is described in \S\ref{sec:axion}. We
discuss some inflation models and isocurvature fluctuations of the
axion in \S\ref{sec:inflation-models}.  \S\ref{sec:conclusion} is
devoted to conclusions and discussion.

Throughout this paper, we set the gravitational scale $\sim 2.4\times
10^{18}$GeV equal to unity.

\section{Comparison with observations}
\label{sec:obs}

The density field of the universe is often described in terms of the
density contrast,
$    \delta(\mbox{\boldmath $x$})\equiv
    \delta \rho(\mbox{\boldmath $x$})/\bar{\rho}
    = \left(\rho(\mbox{\boldmath $x$})-\bar{\rho}\right)/{\bar{\rho}},
$
and its Fourier components
$
    \delta_{k} = \frac1{V}\int d^3\mbox{\boldmath $x$}\ 
    \delta(\mbox{\boldmath $x$}) 
    \exp \left( i\mbox{\boldmath $k$} \cdot \mbox{\boldmath $x$} \right),
$
where $\bar{\rho}$ is the average density of the universe,
$\mbox{\boldmath $x$}$ denotes comoving coordinates, and $V$ is a
sufficiently large volume. If $\delta(\mbox{\boldmath $x$})$ is a
random Gaussian field as predicted by the inflation, then the
statistical properties of the cosmic density field are completely
contained in the matter power spectrum $P(k)\equiv V
\langle|\delta_{k}|^2\rangle$, where $\langle \cdots \rangle$ 
represents ensemble average.

In general, the inflation predicts an adiabatic primordial spectrum,
$P_{\rm ad}^ { prim}(k) \propto k^{n_{s}}$, where $k$ is the
wavenumber of perturbation modes, and $n_{s}$ a spectral index. In
this paper, we set $n_{s}=1$ (the Harrison-Zeldovich spectrum), which
is predicted in a large class of inflation models including hybrid
inflation, new inflation, and the chaotic inflation models discussed
in \S\ref{sec:inflation-models}.  As for isocurvature perturbations,
on the other hand, the primordial spectrum index is conventionally
expressed in terms of entropy perturbations $S_{ar} \equiv \delta_a -
{3\over 4}\delta_r$ as $S_{ar}(k) \propto k^{\tilde{n}_{s}}$, where
$\delta_a$ and $\delta_r$ are density fluctuations of axion and
radiation fields, respectively.  Employing this definition, we can
express the primordial power spectrum as $P_{\rm iso}^{prim}(k)
\propto k^{\tilde{n}_{s}+4}$.  We set ${\tilde n}_{s} =-3 $, which
corresponds to the Harrison-Zeldovich spectrum.

The primordial power spectrum is modified through its evolution in the
expanding universe and the present power spectrum can be described as
\begin{equation}\label{powerspec}
    P(k)= P_{\rm ad} + P_{\rm iso} = A_{\rm ad} k T_{\rm ad}^{2}(k)
    + A_{\rm iso} k T_{\rm iso}^{2}(k),
\end{equation}
where $A_{\rm ad}$ and $A_{\rm iso}$ are normalizations of adiabatic
and isocurvature perturbations, and $T_{\rm ad} (k)$ and $T_{\rm iso}
(k)$ are transfer functions of adiabatic and isocurvature
perturbations.  Conventionally, the transfer function of the
isocurvature perturbations is defined for the initial entropy
perturbations.  Therefore the present matter power spectrum may be
written as $P_{\rm iso} = \tilde{A}_{\rm iso} k^{-3} \tilde{T}^2_{\rm
  iso}(k)$.  For this definition, $\lim_{k \rightarrow \infty}
\tilde{T}_{\rm iso}(k) = 1$, while $\lim_{k \rightarrow 0} {T}_{\rm
  iso}(k) = 1$ for Eq.(\ref{powerspec}).  However, this definition
makes direct comparison between adiabatic and isocurvature matter
power spectra difficult.  Therefore we employ Eq.(\ref{powerspec}) to
define the amplitude and the transfer function for both adiabatic and
isocurvature perturbations.

Here we define the ratio $\alpha$ of the isocurvature to the adiabatic
power spectrum as~\cite{KSY,KKSY}\footnote{This parameter can be
  expressed by using the Hubble parameter during inflation and axion
  breaking scale, as Eq.(\ref{alpha-relation}).}
\begin{equation}\label{alpha-definition}
    \alpha \equiv {A_{\rm iso} \over A_{\rm ad}}~.
\end{equation}
It is well known that the transfer functions for adiabatic CDM models
are essentially controlled by a single parameter, $\Gamma \equiv
\Omega_0h$,~\cite{BBKS} or more precisely, $\Gamma \equiv
\Omega_0h(T_0/2.7K)^{-2}\exp(-\Omega_B-\sqrt{2h}
\Omega_B/\Omega_0)$.~\cite{Sugiyama} Here $h$ is the present Hubble
parameter normalized by $100$km s$^{-1}$ Mpc$^{-1}$, $T_{0}$ is the
present cosmic temperature, and $\Omega_{B}$ is the ratio of the
present baryon density to the critical density. From the galaxy survey
Peacock and Dodds~\cite{Peacock} estimated that $\Gamma\simeq 0.25\pm
0.05+0.32(n_{s}^{-1} -1)$.

In observational cosmology, the quantity $\sigma_8$, which is the
linearly extrapolated rms of the density field in spheres of radius
$8h^{-1}$Mpc, is often used to evaluate amplitudes of the density
fluctuations. This is motivated by the fact that the rms fluctuation
in the number density of bright galaxies measured in a sphere of
radius of $8h^{-1}$Mpc is almost unity.~\cite{Davis} Employing the top
hat window function, we obtain
\begin{equation}
  \sigma^2_8 = \frac1{2\pi^2}\int \frac{dk}k k^3 P(k) \left.\left( \frac{3
      j_1(kr_0)}{kr_0}\right)^2 \right|_{r_{0} =
    8h^{-1}{\rm Mpc}} ~ ,
\end{equation}
where $j_1$ is the spherical Bessel function.

Anisotropies of the cosmic microwave background radiation (CMB),
$\delta T/T$, can be expanded as
\begin{equation}
    \frac{\delta T}{T} (\mbox{\boldmath $\gamma$}) 
    = \sum_{\ell = 2}^{\infty} \sum_{m = -\ell}^{m = +\ell} 
    a_{\ell m} Y_{\ell m}(\mbox{\boldmath $\gamma$}),
\end{equation}
where $Y_{\ell m}$ is a spherical harmonic function, and {\boldmath
  $\gamma$} denotes the direction in the sky. The temperature
autocorrelation function (which compares the temperatures at two
different points in the sky separated by an angle $\theta$) is defined
as
$
    \left\langle{\frac{\delta T}{T}(\mbox{\boldmath $\gamma$}) 
        \frac{\delta T}{T}(\mbox{\boldmath $\gamma'$}) } \right\rangle 
    = \frac{1}{4\pi} \sum _{\ell} (2 {\ell}+1) C_{\ell} P_{\ell}(\cos
    \theta),
    $ where {\boldmath $\gamma$}$\cdot$ {\boldmath $\gamma'$} $= \cos
    \theta$, and $P_l$ is the Legendre polynomial of degree $\ell$.
    The coefficients $C_{\ell}$ are the multipole moments:
\begin{equation}
    (2\ell +1) C_{\ell}=
    \langle \sum_{m=-\ell}^{\ell}\left| a_{\ell m}\right|^2 \rangle.
\end{equation}
The predictions for the CMB anisotropies can be obtained by numerical
integration of the general relativistic Boltzmann equations.

\begin{figure}
    \centerline{\psfig{figure=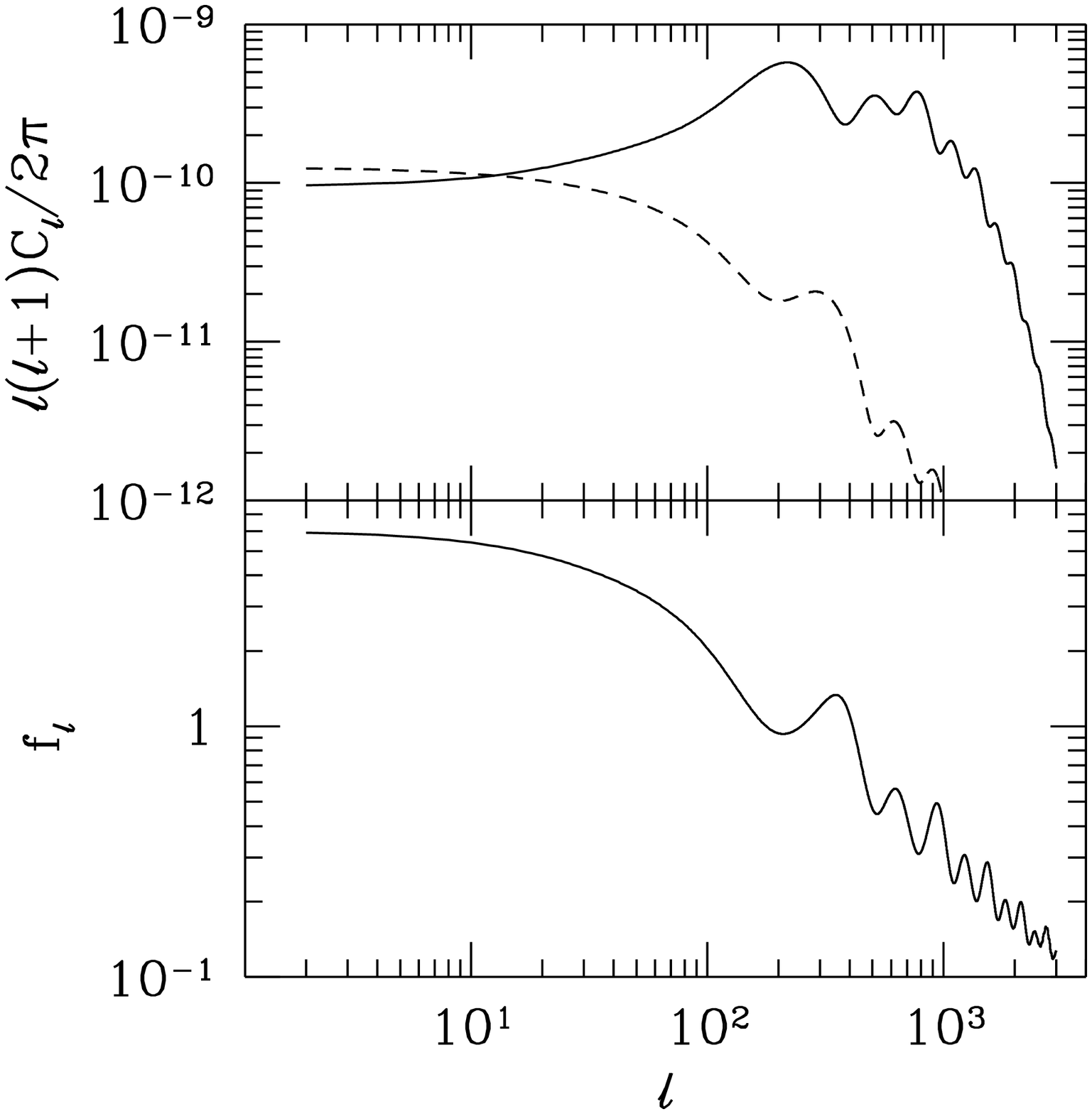,height=15cm} }
    \vspace{1cm}
    \caption{ A sample of the CMB anisotropies angular power spectrum
      (upper panel) and $f_{\ell}$ (lower panel).  In the upper panel,
      the solid line corresponds to pure adiabatic fluctuations, and
      the short dashed line to pure isocurvature fluctuations. We have
      chosen $\Omega_0=1,\ \lambda_0=0,\ h=0.5,\ n=1$ and $\Omega_B
      h^2 = 0.015$ here.}
  \label{fig:cmb-sample}
\end{figure}

Since the isocurvature fluctuations are independent of adiabatic
fluctuations and give different contributions to the CMB anisotropies,
the CMB multipoles can be expressed as a linear combination of two
components as
\begin{equation}\label{pre-csubell-alpha}
    C_\ell(\alpha) = g(\alpha) C^{\rm ad}_\ell + 
    h(\alpha)C^{\rm iso}_\ell ~ ,
\end{equation}
where $C^{\rm ad}_\ell$ and $C^{\rm iso}_\ell$ are adiabatic and
isocurvature components of CMB multipoles, respectively, and
$g(\alpha)$ and $h(\alpha)$ are functions of $\alpha$.  Note that
$C^{\rm ad}_\ell = C_\ell(0)$ and $C^{\rm iso}_\ell = C_\ell(\infty)$.
Employing the COBE normalization obtained from the 4-yr
data,~\cite{COBE-norm} which fixes $C_{10}$, i.e., $C_{10}=C^{\rm
  ad}_{10}=C^{\rm iso}_{10}$, we obtain $h(\alpha)=1-g(\alpha)$.  Then
we introduce a relative ratio between the isocurvature and adiabatic
components at $\alpha=1$ as $f_\ell^2 \equiv \left(1-g(1)\right)C^{\rm
  iso}_\ell/g(1)C^{\rm ad}_\ell$.  In the large scale limit ($\ell
\rightarrow 2$), this factor is almost $f_\ell\approx 6$, which is
expected from the differences between Sachs-Wolfe contributions of
adiabatic and isocurvature fluctuations.~\cite{KSBEHS} On the COBE
scale ($\ell\simeq 10$), however, $f_\ell$ is smaller than 6.
Furthermore, for larger $\ell$, $f_\ell$ strongly depends on $\ell$
because the angular power spectra for the adiabatic and the
isocurvature fluctuations differ greatly (see
Fig.\ref{fig:cmb-sample}).  The precise value of $f_{10}$ depends on
$\Omega_0$ and $h$.  For example, $f_{10}^2\simeq 30$ for $\Omega_0=1$
and $h\gtrsim 0.6$, and it becomes smaller when $\Omega_0$ or $h$
becomes smaller.  Using this factor $f_\ell$, we can write the ratio
of the isocurvature to the adiabatic mode at arbitrary $\alpha$ as
\begin{equation}
f_\ell^2\alpha = {\left(1-g(\alpha)\right)C^{\rm iso}_\ell \over
  g(\alpha)C^{\rm ad}_\ell}~ .
\end{equation}
Using the COBE normalization, we get $g(\alpha)=
1/(1+f_{10}^2\alpha)$.  Eventually, if we know the value of $f_{10}$,
the COBE normalized pure adiabatic ($C^{\rm ad}_\ell$) perturbations,
and isocurvature ($C^{\rm iso}_\ell$) perturbations, we can obtain the
COBE normalization for any admixture of adiabatic and isocurvature
perturbations as
\begin{equation}
    \label{csubell-alpha}
    C_\ell(\alpha) = {C^{\rm ad}_\ell + f_{10}^2\alpha C^{\rm iso}_\ell
    \over   {1 + f_{10}^2\alpha}} ~.
\end{equation}

A similar argument can be applied to $\sigma_8$ as 
\begin{equation}
    \label{sigma8-alpha}
    \sigma_8(\alpha) =  \sqrt{\frac{(\sigma^{\rm ad}_8)^2
        +  f_{10}^2 \alpha (\sigma^{\rm iso}_8)^2} {1 +
        f_{10}^2 \alpha}},
\end{equation}
where $\sigma_8^{\rm ad}$ and $\sigma_8^{\rm iso}$ are the values of
$\sigma_8$ normalized to COBE for the cases of pure adiabatic and pure
isocurvature perturbations, respectively.

We can obtain $\sigma_{8}$ as a function of $\Omega_{0}$ from
observations of the cluster abundance, since this abundance is very
sensitive to the amplitude of the density fluctuations.  Here we adopt
the values of $\sigma_8$ which are obtained from the analysis of the
local cluster X-ray temperature function:~\cite{Eke}
\begin{equation}
    \label{sigma8-obs}
  \sigma_8 = (0.52\pm0.04)\Omega_0^{-0.52+0.13\Omega_0}\quad ({\rm
    for}\quad \Omega_{0} + \lambda_{0} = 1) .
\end{equation}
In other analyses,~\cite{White,Viana,Kitayama} similar values for
$\sigma_8$ have been obtained.

\begin{figure}
    \centerline{\psfig{figure=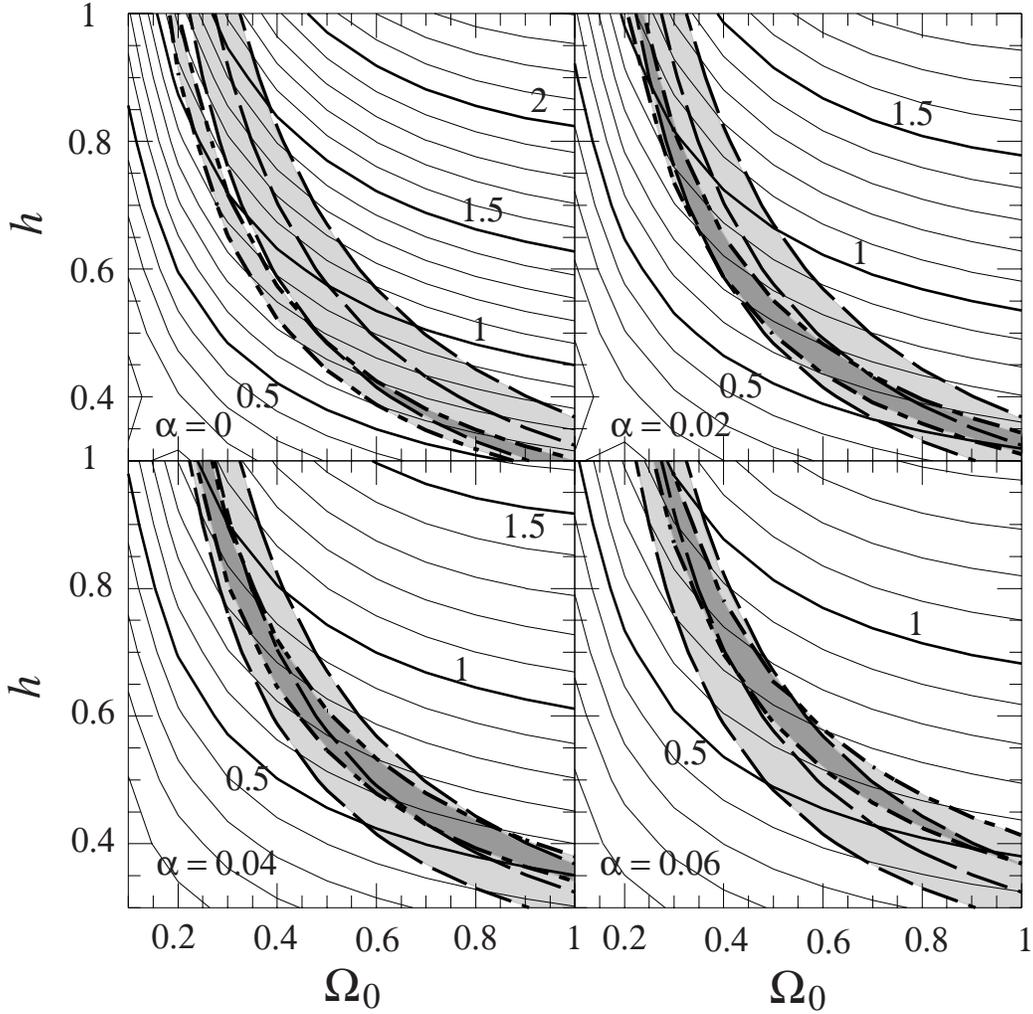,height=15cm}}
    \vspace{1cm}
    \caption{$\sigma_8(\alpha)$ as functions of $\Omega_0$ and $h$
      normalized by COBE. The four panels correspond to $\alpha = 0$
      (top left), $0.02$ (top right), $0.04$ (bottom left) and $0.06$
      (bottom right). The lightly shaded region inside the two short
      dash - long dash lines indicates the values of $\sigma_8$
      observed from the X-ray clusters of the galaxies.  The three
      long dashed lines represent shape parameters $\Gamma = 0.2,
      0.25$ and $0.3$ from left to right, and the observationally
      favored region ($\Gamma = 0.25\pm 0.05$) is also lightly shaded.
      The darkly shaded region satisfies the constraints from both the
      cluster abundances and the shape parameter.}
    \label{fig:sigma8_1} 
\end{figure}

\begin{figure}
    \centerline{\psfig{figure=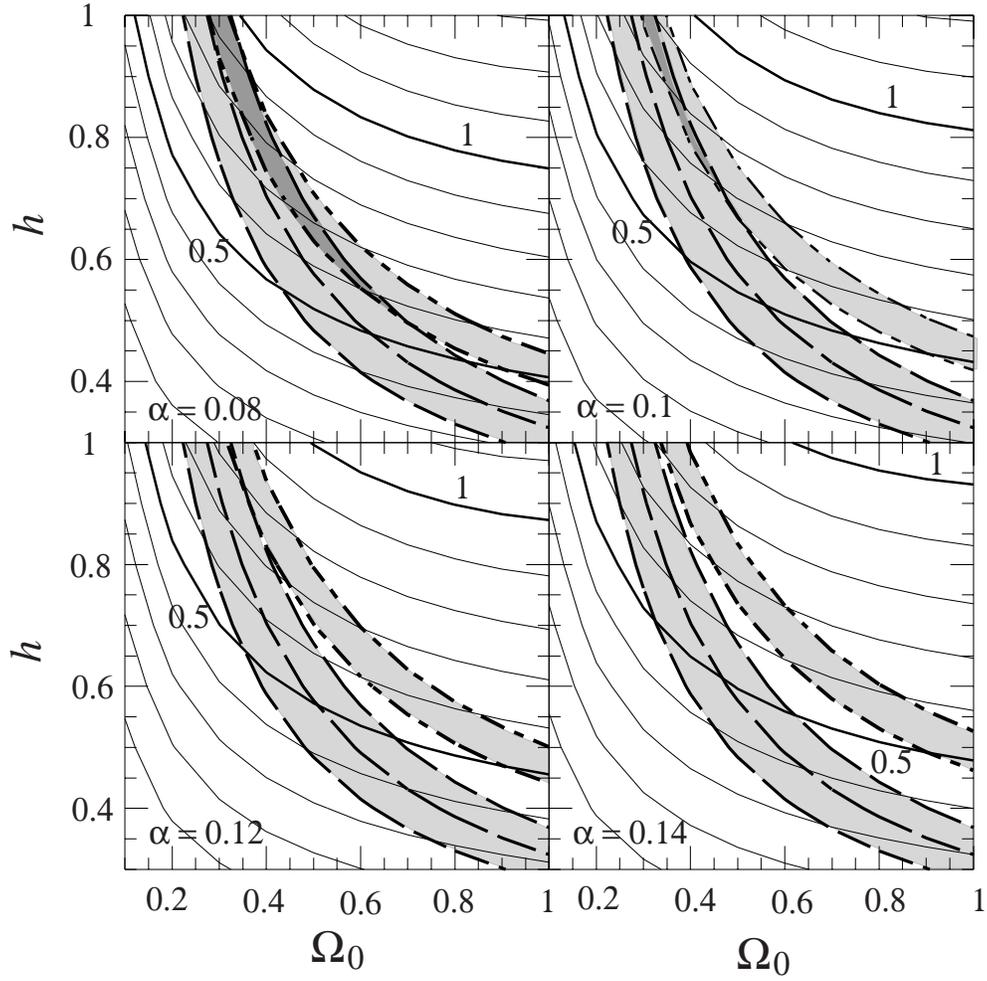,height=15cm}}
    \vspace{1cm}
    \caption{The same as Fig.~\protect\ref{fig:sigma8_1}, except that the
      four panels correspond to $\alpha = 0.08$ (top left), $0.10$
      (top right), $0.12$ (bottom left) and $0.14$ (bottom right).}
    \label{fig:sigma8_2} 
\end{figure}

\begin{figure}
    \centerline{\psfig{figure=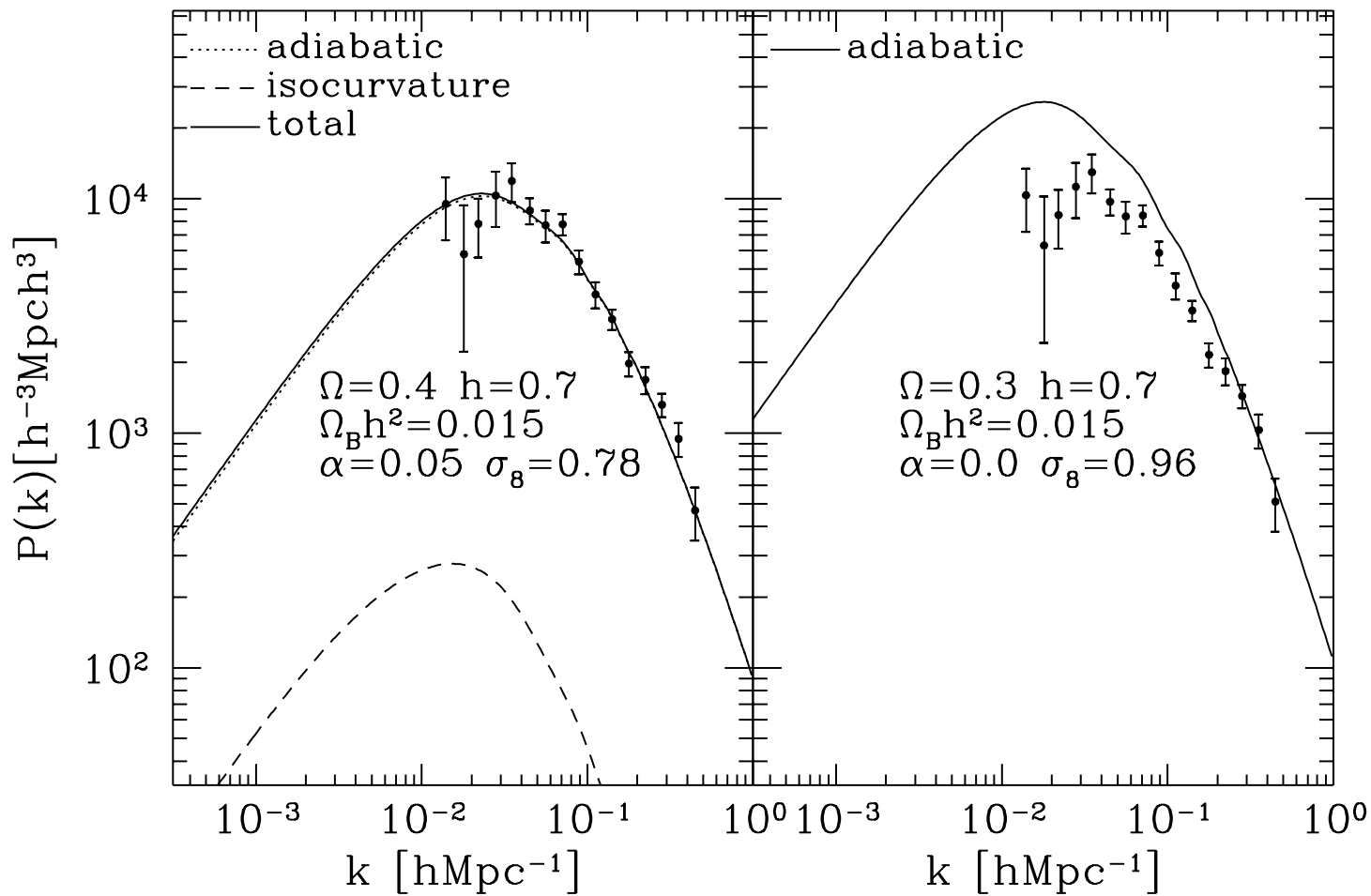,height=15cm,angle=-90}}
    \vspace{1cm}
    \caption{The power spectra for $\alpha = 0.05, \Omega_0=0.4$ and
      $h=0.7$ (left panel) and $\alpha = 0, \Omega_0=0.3$ and $h=0.7$
      (right panel). The symbols denote the observational
      data.~\protect\cite{Peacock}}
    \label{fig:power-spec}
\end{figure}

In Figs.\ref{fig:sigma8_1} and \ref{fig:sigma8_2}, we plot
$\sigma_8(\alpha)$ as functions of $\Omega_0$ and $h$. The observed
$\sigma_8$ mentioned above is also drawn as a region inside the two
short dash - long dash lines. Furthermore, three dashed lines,
corresponding to the shape parameters $\Gamma = 0.2, 0.25$, and $0.3$
from left to right, are drawn in the figures. From these figures, one
can see that, for example, if $h\simeq 0.7$ and $\Gamma=0.25\pm 0.05$,
the observed $\sigma_8$ is inconsistent with the COBE normalized
predictions for $\alpha=0$. It is possible, of course, to fit the
observation with $\alpha = 0$ if $h\simeq 0.4 (\Omega_{0} \simeq
0.6)$. However, small values of the Hubble parameter ($h\simeq 0.4$)
seem unlikely from the recent observations,~\cite{Freedman} which
suggest $h = 0.73\pm 0.06({\rm stat}) \pm 0.08({\rm sys})$.  Rather,
$\alpha=0.04$ to $0.06$ give better fits for $h\simeq 0.7$.  This fact
is also seen in Fig.\ref{fig:power-spec} (left panel), where the power
spectrum for $\alpha = 0.05, \Omega_0=0.4$ and $h=0.7$ is shown
together with observational data.~\cite{Peacock} In
Fig.\ref{fig:power-spec} we also plot the pure adiabatic power
spectrum for $\Omega_0=0.3$ and $h=0.7$ for comparison (right panel).
The values of $\Omega_0$ for both power spectra are chosen to satisfy
Eq.~(\ref{sigma8-obs}) for $h=0.7$. It is clear that the mixed model
($\alpha \simeq 0.05$) gives a much better fit to the data obtained
from the galaxy survey.

\begin{figure}
\centerline{\psfig{figure=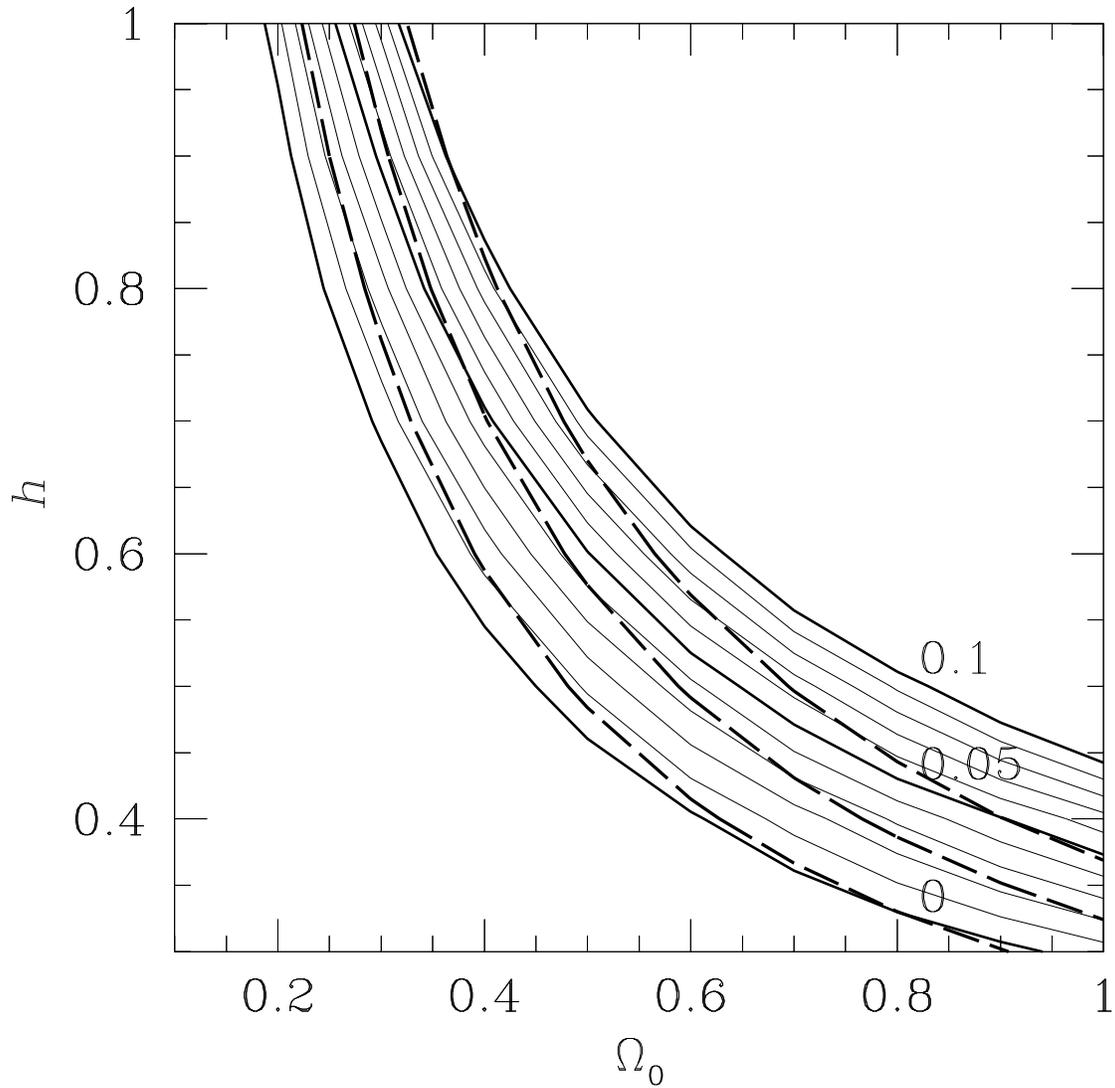,height=15cm}}
    \vspace{1cm}
    \caption{The solid lines represent values of $\alpha$ which give the best
      fit to the data of $\sigma_8$ from cluster abundances. The three
      long dashed lines represent shape parameters $\Gamma = 0.2,
      0.25$ and $0.3$ from left to right.}
    \label{fig:alpha-best}
\end{figure}

For each $\Omega_0$ and $h$, we can obtain the best fit values of
$\alpha$, which are plotted in Fig.\ref{fig:alpha-best}. The three
dashed lines correspond to $\Gamma = 0.2, 0.25$, and $0.3$, from left
to right. One can see from this figure, for example, for $h=0.7$ and
$\Gamma=0.25$, the best fit value of $\alpha$ to observations is
$\alpha \simeq 0.05$. This figure also indicates that if
$\Gamma=0.25\pm 0.05$, the observations of $\sigma_8$ require
$\alpha\lesssim 0.1$. This is also seen in Fig.\ref{fig:sigma8_2}.

In the above discussion, we see that $\alpha\sim {\cal O}(10^{-2})$ is
consistent with observations of large scale structures. Let us examine
what this value predicts. As mentioned earlier, the isocurvature
fluctuations give different contributions to the CMB anisotropies than
the adiabatic fluctuations. If one takes very large values of
$\alpha$, the prominent acoustic peaks disappear, and only the
Sachs-Wolfe plateau exists in the angular power spectrum.  Therefore,
we must investigate the height of the acoustic peak(s) predicted by
our model.

\begin{figure}
\centerline{\psfig{figure=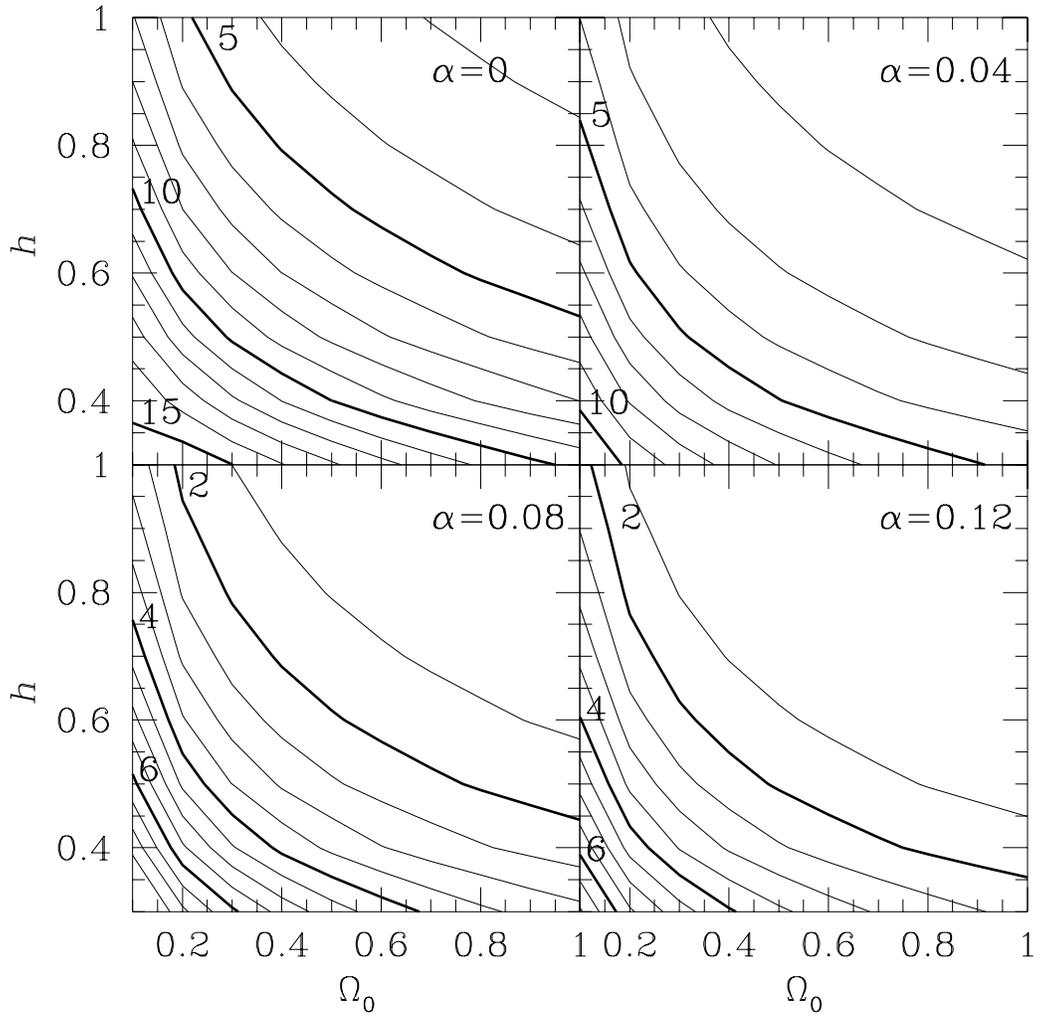,height=15cm}}
    \vspace{1cm}
    \caption{Peak-height parameter $\gamma$ (see
      Eq.(\protect\ref{peak-height})) for $\alpha = 0$ (top left),
      $0.04$ (top right), $0.08$ (bottom left) and $0.12$ (bottom
      right).}
    \label{fig:peakheights}
\end{figure}

In order to predict this height quantitatively, we introduce the
peak-height parameter in the CMB anisotropy angular power spectrum
defined by
\begin{equation}
    \label{peak-height}
    \gamma\equiv \frac{\left. \ell (\ell + 1) C_\ell \right |_{\ell =
    \ell_{\rm peak}}}{\left. \ell (\ell + 1) C_\ell \right |_{\ell =
    10}},
\end{equation}
where $\gamma,\ C_\ell$ and $\ell_{\rm peak}$ are all functions of
$\alpha$.  Here $\left. \ell (\ell + 1) C_\ell \right |_{\ell =
  \ell_{\rm peak}}$ denotes the highest value, which is usually the
value at the first acoustic peak.  When $\alpha$ becomes larger, the
height of the acoustic peak becomes lower. If the peak height becomes
lower than the Sachs-Wolfe plateau, this peak-height parameter tends
to be unity. In Fig.\ref{fig:peakheights}, we plot this quantity.  One
can see that, for example, for $\alpha \simeq 0.05$, $\Omega_{0}=0.4$,
and $h=0.7$, this ratio is about $\gamma\simeq 2$ - $3$.

\begin{figure}
\centerline{\psfig{figure=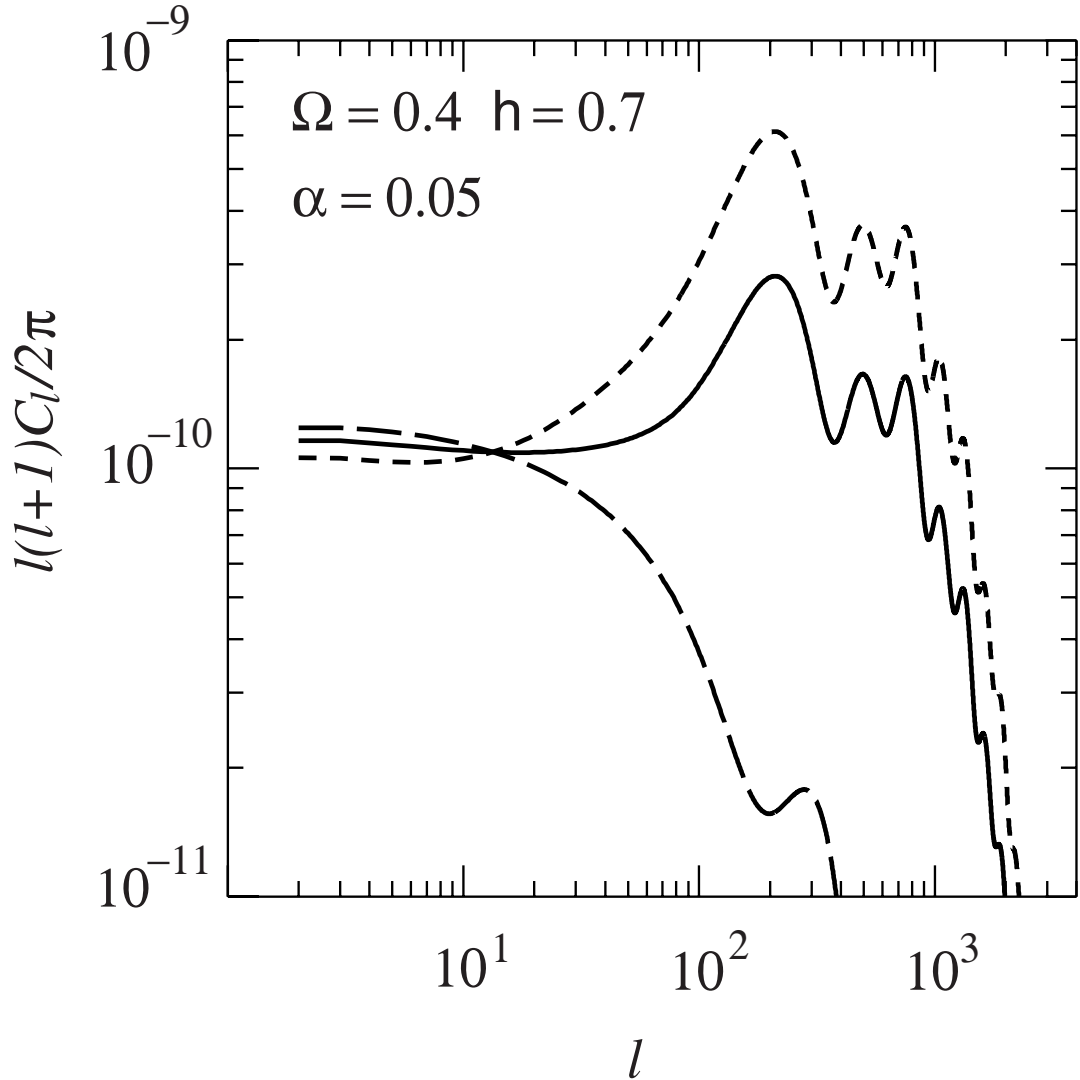,height=15cm} }
    \vspace{1cm}
    \caption{ The CMB angular power spectra normalized by COBE.
      Here we have chosen $\Omega_0=0.4,\ h=0.7,\ \lambda_0=0.6$, and
      $\Omega_B h^2=0.015$.  The short dashed line corresponds to the
      pure adiabatic case ($\alpha=0$), the long dashed line to the
      pure isocurvature case ($\alpha=\infty$), and the solid line to
      the case $\alpha = 0.05$, which gives the best fit to the
      observational data.}
  \label{fig:mix_sample}
\end{figure}

\begin{figure}
\centerline{\psfig{figure=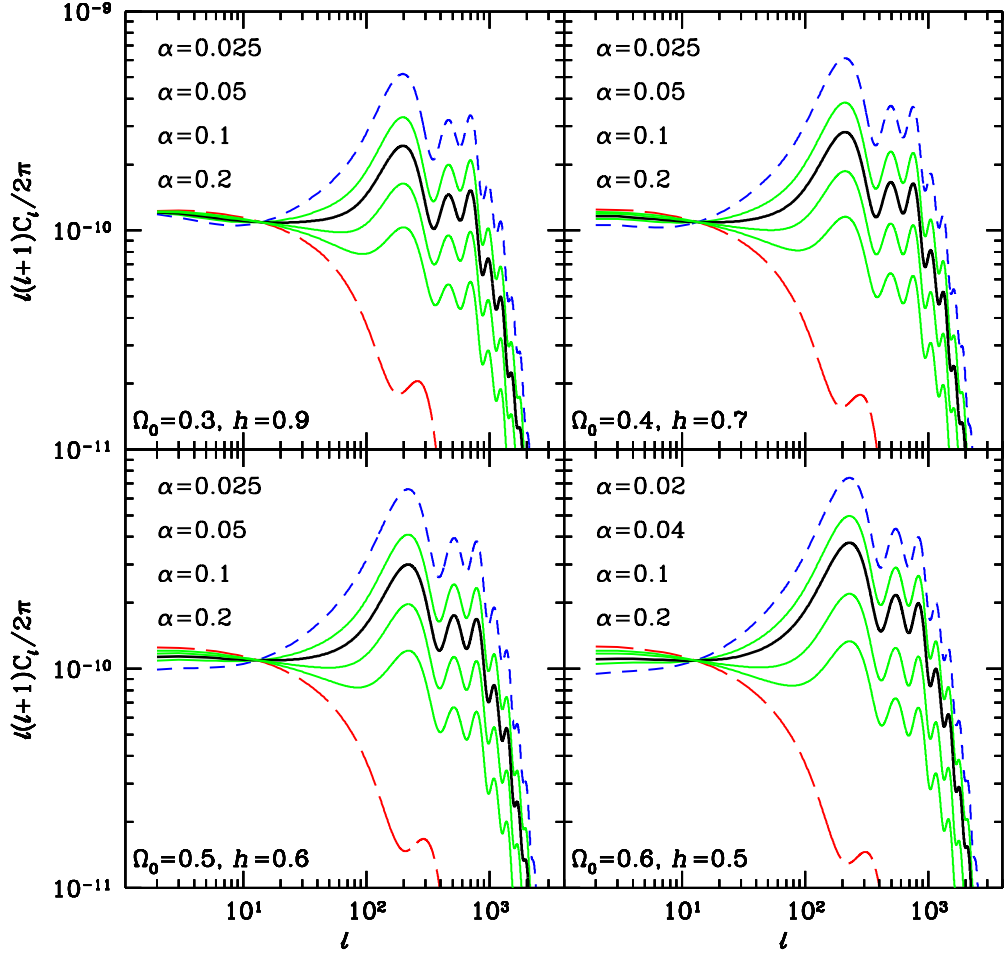,height=15cm} }
    \vspace{1cm}
    \caption{The same as Fig.~\protect\ref{fig:mix_sample}, except
      that the cosmological parameters $\Omega_0$ and $h$ have been
      changed: $\Omega_0=0.3,\ h=0.9$ (top left), $\Omega_0=0.4,\ 
      h=0.7$ (top right), $\Omega_0=0.5,\ h=0.6$ (bottom left) and
      $\Omega_0=0.6,\ h=0.5$ (bottom right).  In this figure, we also
      plot angular power spectra for various $\alpha$ to see how much
      the acoustic peak decrease as $\alpha$ changes.}
  \label{fig:mix_multi}
\end{figure}

In Fig.\ref{fig:mix_sample}, we plot the CMB angular power spectra
normalized by the COBE for the cases $\alpha = 0,\ 0.05$, and
$\infty$. Also, in Fig.\ref{fig:mix_multi}, we plot the same angular
power spectra, changing the cosmological parameters $\Omega_0$ and
$h$.  From these figures, one can see how much the height of the
acoustic peak decreases when $\alpha$ is large.

Future satellite experiments such as MAP~\cite{MAP} and
PLANCK~\cite{PLANCK} are expected to measure the CMB anisotropies with
fine angular resolutions and detect the peak-height parameter.  If
this is done, our model can be tested by these observations.

\section{Fluctuations of axions}
\label{sec:axion}

During the inflation, the axion experiences quantum fluctuations whose 
amplitude is given by 
\begin{equation}
    (\delta a (k))^2 = \frac{H^2}{2k^3},
\end{equation}
where $k$ is the comoving wavenumber and $H$ the Hubble parameter
during the inflation.  These fluctuations become classical due to the
exponential expansion of the universe.  After the axion acquires a
mass at the QCD scale, the axion fluctuations lead to density
fluctuations. The density fluctuations of axions are of an
isocurvature type, because they are massless during the inflation and
hence do not contribute to the fluctuations of the total density of
the universe. The density of axions is written as
\begin{equation}
    \rho_a = \frac{1}{2}m_a^2 a^2 = \frac{1}{2}m_a^2F_a^2\theta_a^2,
\end{equation}
where $m_a$ is the mass of the axion and $\theta_a$ the phase of the
Peccei-Quinn scalar ($0 \le \theta_a < 2\pi$).  Then the isocurvature
fluctuations with comoving wavenumber $k$ are given by
\begin{equation}
    \label{iso-fluc}
    \delta_a^{\rm iso}(k) \equiv 
    \left(\frac{\delta \rho_a}{\rho_a}(k)\right)_{\rm iso} 
    = \frac{ 2\delta a}{a} = \frac{\sqrt{2} H}{F_a \theta_a}k^{-3/2},
\end{equation}
where we redefine $H$ as the Hubble parameter when the comoving
wavelength $k^{-1}$ becomes equal to the Hubble radius $H^{-1}$ during
the inflation epoch. It should be noted that the above initial
spectrum is of the Harrison-Zeldovich type in the case of isocurvature
perturbations (see \S\ref{sec:obs}).

On the other hand, the inflaton itself generates adiabatic
fluctuations given by
\begin{equation}
    \label{ad-fluc}
    \delta_a^{\rm ad}(k) \equiv \left(\frac{\delta
        \rho}{\rho}(k)\right)_{\rm ad} = \frac{2\sqrt{2} H^3}{3 V'
      \tilde{H}^2 R(t)^2}k^{1/2},
\end{equation}
where $V$ is the potential for an inflaton, and $R(t)$ and $\tilde{H}$
are the scale factor and Hubble constant at some arbitrary time
$t$.~\cite{MFP} To compare these two types of fluctuations, it may be
natural to consider the ratio of the power spectra at horizon
crossing, i.e.  $k^{-1} R = \tilde{H}^{-1}$, which is written
as~\cite{KSY}\footnote{Our Eq.(\ref{ratio}) is different from Eq.(5)
  in Ref. 1) by a factor $1/4$ due to a typo.}
\begin{equation}
    \label{ratio}
    \alpha_{\rm KSY} = \left. 
      \frac{P_{\rm iso}}{P_{\rm ad}} 
    \right|_{k/R=\tilde{H}} =
    \frac{9(V')^2}{4H^4 F_a^2 \theta_a^2}.
\end{equation}
However, this definition is different from the ratio $\alpha$ of the
{\it present} power spectra in Eq.(\ref{alpha-definition}) since we
need to take into account the difference between the time evolution of
adiabatic and isocurvature perturbations for the radiation dominant
regime and the matter dominant epoch.~\cite{KSBEHS} Eventually, we can
obtain the following relation:~\cite{KKSY}\footnote{ The factor
  $(2/15)$ comes from the value of the transfer function in the long
  wavelength limit,~\cite{Kodama-Sasaki} and the extra factor $(10/9)$
  is due to the decay of the gravitational potential at the transition
  from the radiation dominated universe to the matter dominated
  universe. }
\begin{equation}\label{alpha-relation}
  \alpha=
  \left( \frac{2}{15} \right)^2 \left( \frac{10}{9} \right)^2
  \alpha_{\rm KSY} = \left( \frac{4}{27} \right)^2 \alpha_{\rm KSY} =
  \frac{4(V')^2}{81H^4F_a^2\theta_a^2} .
\end{equation}
In this notation, $\alpha=1$ means that the adiabatic and the
isocurvature matter power spectra become the same in the long
wavelength limit (see Eq.(\ref{alpha-definition})).

As mentioned in the previous section, the isocurvature fluctuations
give contributions to the COBE measurements which are larger than the
adiabatic fluctuations by the factor $f_{10}$ (if we take
$\Omega_0\simeq 1$ and $h\simeq 0.6$, this factor is approximately
$\approx \sqrt{30}$).  Therefore, when we take account of the
isocurvature fluctuations, the correct COBE normalization is expressed
as~\cite{COBE-norm}
\begin{equation}\label{iso-cobe}
  \left. \frac{V^{3/2}}{V'}\right|_{N=60}\simeq \frac{5.3\times
    10^{-4}}{\sqrt{1+f_{10}^2\alpha}}.
\end{equation}
Here, we have ignored the tensor perturbations.

From Eqs.~(\ref{alpha-relation}) and (\ref{iso-cobe}), we see that
\begin{equation}\label{fa-alpha}
  \frac{1 + f_{10}^2\alpha}{\alpha} \simeq 21\times \left(
    \frac{H}{10^{12}{\rm GeV}} \right)^{-2} \left( \frac{F_a
      \theta_a}{10^{16}{\rm GeV}} \right)^{2}.
\end{equation}
Thus, if we know the value of $\alpha$ and $H$,
we have a constraint on $F_{a}$.

\section{Inflation models}
\label{sec:inflation-models}

In order to determine the size of isocurvature fluctuations produced
during inflation, here we consider some inflation models.

There are mainly three classes of realistic inflation models: hybrid
inflation models, new inflation models, and chaotic inflation models.
We adopt simple realizations of the former two types of models in the
context of supergravity.  Although it is difficult to achieve chaotic
inflation in supergravity, we also consider the chaotic type for
comparison, since it is very simple and is often considered in the
literature.

In all cases, we have a reasonable parameter region which produces
observable isocurvature fluctuations.

\subsection{Hybrid inflation model}
\label{subsec:hybrid}

In this subsection, we consider the hybrid inflation model proposed by
Linde and Riotto,~\cite{Linde-Riotto} because this hybrid inflation
takes place under natural initial conditions and is consistent with
supergravity (SUGRA)~\cite{Wess-Bagger} (see also Refs. 23) and 24)).
Let us now consider the hybrid inflation model,~\cite{Linde-Riotto}
which contains two kinds of superfields: $S(x,\ \theta)$ and $\psi(x,\ 
\theta)$ together with $\bar{\psi}(x,\ \theta)$. Here $\theta$ is the
Grassmann number denoting superspace.~\cite{Wess-Bagger} The model is
based on $U(1)_R$ symmetry under which $S(\theta)\to
e^{2i\alpha}S(\theta e^{-i\alpha})$ and $\psi(\theta)
\bar{\psi}(\theta)\to \psi(\theta e^{-i\alpha}) \bar{\psi}(\theta
e^{-i\alpha})$. The superpotential is then given by
\begin{equation}
    W=S(-\mu^2+\kappa \bar{\psi}\psi),
\end{equation}
where $\mu$ is a mass scale and $\kappa$ a coupling constant. The
scalar potential obtained from this superpotential, in the global SUSY
limit, is
\begin{equation}
  V = \left|-\mu^2+\kappa \bar{\psi} \psi \right|^2+\kappa^2\left| S
  \right|^2 \left(\left|\psi \right|^2+ \left|\bar{\psi}\right|^2
  \right)+D{\rm - terms},
\end{equation}
where scalar components of the superfields are denoted by the same
symbols as the corresponding superfields. The potential minimum,
\begin{equation}
  \langle S \rangle=0,\ \langle \psi \rangle
  \langle\bar{\psi}\rangle=\frac{\mu^2}\kappa,\ \left|
    \langle\psi\rangle \right|=\left| \langle \bar{\psi} \rangle
  \right|,
\end{equation}
lies in the $D$-flat direction $\left|\psi \right| = |
\bar{\psi}|$.\footnote{
  We have assumed a $U(1)$ gauge symmetry, where $\psi(x,\ \theta)$
  and $\bar{\psi}(x,\ \theta)$ have opposite charges of the $U(1)$, so
  that the $\psi\bar{\psi}$ term is allowed.}
By the appropriate gauge and $R$-transformations in this $D$-flat
direction, we can bring the complex $S,\ \psi$ and $\bar{\psi}$ fields
onto the real axis:
\begin{equation}
  S\equiv\frac{1}{\sqrt{2}}\sigma,\ \psi=\bar{\psi}\equiv\frac12\phi,
\end{equation}
where $\sigma$ and $\phi$ are canonically normalized real scalar
fields. The potential in the $D$-flat directions then becomes
\begin{equation}\label{D-flat-potential}
  V(\sigma,\ \phi) = \left(-\mu^2 + \frac14\kappa\phi^2\right)^2 +
  \frac14 \kappa^2\sigma^2\phi^2,
\end{equation}
and the absolute potential minimum appears at $\sigma=0,\ 
\phi=\bar{\phi}=\mu/\sqrt{\kappa}$. However, for $\sigma > \sigma_c
\equiv \sqrt{2}\mu/\sqrt{\kappa}$, the potential has a minimum at
$\phi=0$. The potential Eq.(\ref{D-flat-potential}) for $\phi = 0$ is
exactly flat in the $\sigma$-direction. The one-loop corrected
effective potential (along the inflationary trajectory $\sigma >
\sigma_c$ with $\phi = 0$) is given by~\cite{Coleman-Weinberg,Dvali}
\begin{eqnarray}\label{one-loop-correction}
  V_{\rm one-loop}&=& \frac{\kappa^2}{128 \pi^2} \left[ (\kappa
    \sigma^2 - 2\mu^2)^2 \ln{\frac{\kappa \sigma^2 -
        2\mu^2}{\Lambda^2}}\right.\nonumber\\& +&\left.  (\kappa
    \sigma^2 + 2\mu^2)^2 \ln\frac{\kappa \sigma^2 + 2\mu^2}{\Lambda^2}
    - 2 \kappa^2\sigma^4 \ln {\frac{\kappa \sigma^2}{\Lambda^2}}
  \right],
\end{eqnarray}
where $\Lambda$ indicates the renormalization scale.

Next, let us consider the supergravity (SUGRA) effects on the scalar
potential (ignoring the one-loop corrections calculated above). The
$R$-invariant K\"ahler potential is given by~\cite{Panagiotakopoulos}
\begin{equation}
  K(S, \psi,\bar{\psi}) = \left| S \right|^2 + \left| \psi \right|^2 +
  \left| \bar{\psi} \right|^2-\frac\beta4 \left| S \right|^4 +\cdots,
\end{equation}
where the ellipsis denotes higher order terms, which we neglect in the
present analysis. Then, the scalar potential
becomes~\cite{Wess-Bagger}
\begin{eqnarray}
  V(\sigma,\ \phi)&=&\exp\left( \frac{\sigma^2}2 -
    \frac\beta{16}\sigma^4 + \frac{\phi^2}2 \right)
  \left[\frac14\kappa^2 \phi^2 \sigma^2\left (
      1+\frac{\phi^2}4-\frac{\mu^2}{\kappa}\right)\right.\nonumber\\ 
  &&+ \left.\left( 1+ \frac{\beta-1}2 \sigma^2 + \frac{\beta^2 + \beta
        + 1}4 \sigma^4 \right) \left( -\mu^2+\frac\kappa 4
      \phi^2\right)^2\right].
\end{eqnarray}
As in the global SUSY case, for $\sigma\gtrsim\sigma_c$ the potential
has a minimum at $\phi = 0$. The scalar potential for
$\sigma\gtrsim\sigma_c$ and $\phi=0$ becomes
\begin{equation}\label{sugra-correction}
  V_{\rm SUGRA} = \mu^4\left(1+\frac\beta2\sigma^2+\frac{4\beta^2 +
      7\beta + 2}{16}\sigma^4 + \cdots\right).
\end{equation}

In the first approximation, we assume that the inflaton potential for
$\sigma\gtrsim \sigma_c$ and $\phi=0$ is given by the simple sum of
the one-loop corrections Eq.(\ref{one-loop-correction}) and the SUGRA
potential Eq.(\ref{sugra-correction}):
\begin{eqnarray}
  V(\sigma)&=&\mu^4\left(1+\frac\beta2\sigma^2+\frac{4\beta^2 + 7\beta
      + 2}{16}\sigma^4 \right)\nonumber\\ 
  &+& \frac{\kappa^2}{128 \pi^2} \left[ (\kappa \sigma^2 - 2\mu^2)^2
    \ln{\frac{\kappa \sigma^2 - 2\mu^2}{\Lambda^2}}\right.\nonumber\\ 
  &+&\left.  (\kappa \sigma^2 + 2\mu^2)^2 \ln\frac{\kappa \sigma^2 +
      2\mu^2}{\Lambda^2} - 2 \kappa^2\sigma^4 \ln {\frac{\kappa
        \sigma^2}{\Lambda^2}} \right].
\end{eqnarray}
Hereafter, we study the dynamics of the hybrid inflation with this
potential.

We suppose that the inflaton $\sigma$ is chaotically distributed in
space at the Planck time and it happens in some region in space that
$\sigma$ is approximately equal to the gravitational scale and $\phi$
is very small ($\approx 0$). Then, the inflaton $\sigma$ rolls slowly
down the potential, and this region inflates and dominates the
universe eventually. During the inflation, the potential is almost
constant, and the Hubble parameter is given by $H =
V^{1/2}/\sqrt3\simeq\mu^2/\sqrt3$. When $\sigma$ reaches the critical
value $\sigma_c$, the phase transition takes place and the inflation
ends.  In order to solve the flatness and horizon problem we need an
e-folding number $N$ $\simeq 60$.~\cite{Kolb-Turner} In addition, the
adiabatic density fluctuations during the inflation should account for
the observation by COBE, which leads to Eq.~(\ref{iso-cobe}).

The evolutions for $\sigma$ and $N$ are described by 
\begin{eqnarray}
  \dot{\sigma} & = & -\frac{V'}{3H}, \\
  \dot{N} &= & -H,
\end{eqnarray}
which are numerically integrated with $\sigma|_{N=0}=\sigma_c$ and
Eq.(\ref{iso-cobe}) as boundary conditions.  Though the inflaton
potential is parameterized by three parameters (i.e., $\kappa,\ \beta$,
and $\mu$), we can reduce the number of free parameters from three
to two by using the constraint Eq.(\ref{iso-cobe}). We consider $\mu$
as a function of $\kappa$ and $\beta$, i.e.  $\mu = \mu(\kappa,\ 
\beta)$.

When $\sigma \gtrsim 1$, the slow roll approximation cannot be
maintained. Therefore, if the obtained value of $\sigma|_{N=60}\equiv
\sigma_0$ is larger than the gravitational scale, we should
discard those parameter regions.  Also, one of attractions of the
hybrid inflation model is that one does not have to invoke extremely
small coupling constants. Thus we assume all of the coupling constants
$\kappa$ and $\beta$ to have values of ${\cal O}(1)$. Here, we choose
$10^{-2}\lesssim \kappa,\ \beta \lesssim 10^{-1}$ as a ``reasonable
parameter region'' (when $\kappa,\ \beta \gtrsim 10^{-1}$, $\sigma_0$
exceeds unity, and $N$ cannot be as large as 60).  

\begin{figure}
\centerline{\psfig{figure=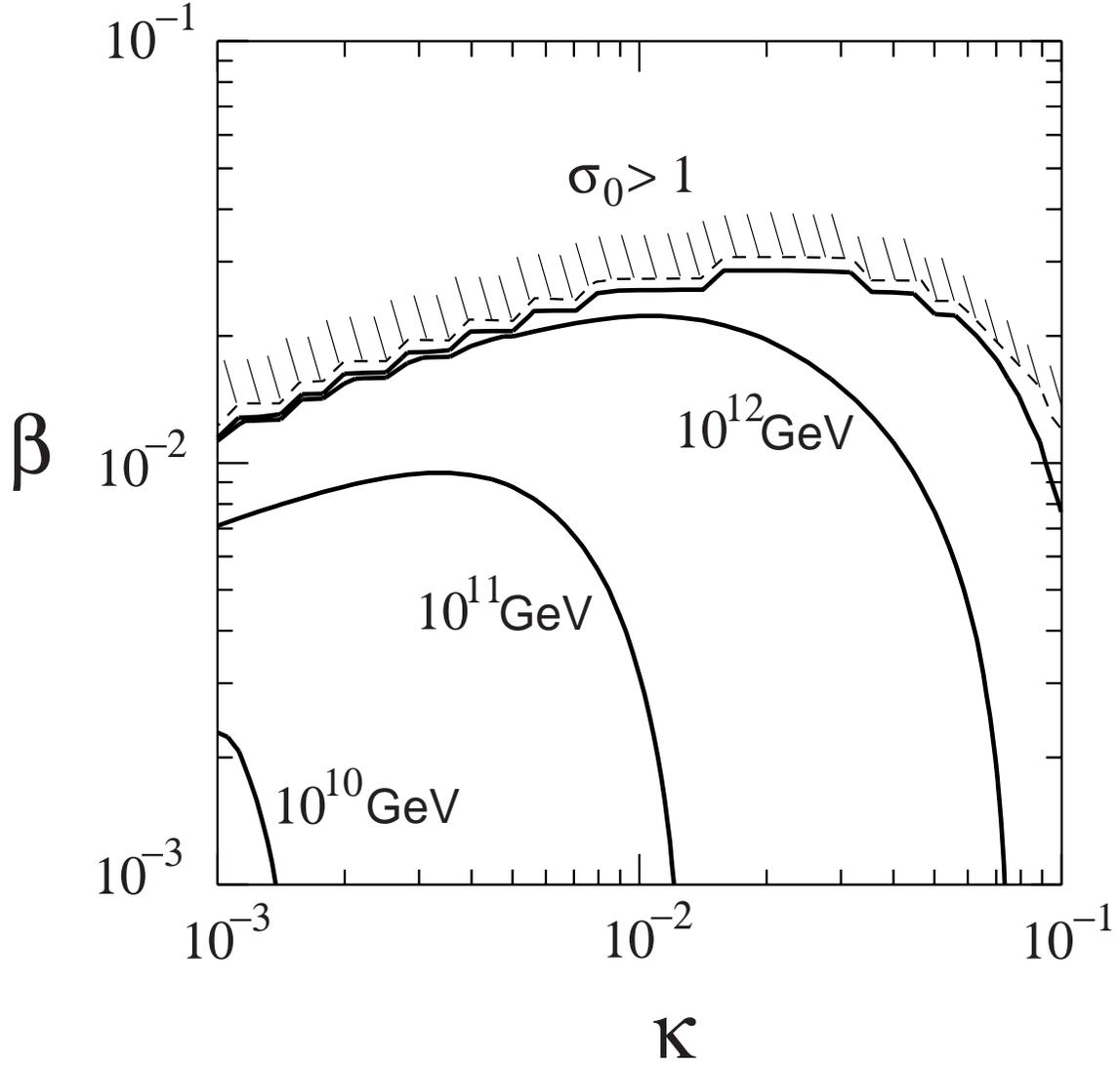,height=15cm} }
    \vspace{1cm}
    \caption{The Hubble parameter during the hybrid inflation, which is
      normalized by the COBE, ignoring tensor perturbations and
      isocurvature fluctuations. In the region above the dashed line,
      $\sigma_0$ exceeds the gravitational scale (i.e., $\sigma_0> 1$)
      and it is excluded.}
  \label{fig:hubble_inflation}
\end{figure}

The result is that $H$ is as large as $H\sim {\cal O}(10^{11-12}{\rm
  GeV})$, with the reasonable values of the coupling constants
$10^{-2}\lesssim \kappa,\ \beta \lesssim 10^{-1}$~\cite{KKSY} (see
Fig.\ref{fig:hubble_inflation}). When $H$ has such a large value, the
inflation generates large isocurvature fluctuations of the axion, if
it exists.

We are at the point to evaluate $\alpha$, which depends on two
parameters $F_a\theta_a$ and $H$, as seen from Eq.(\ref{fa-alpha}). We
take $H\simeq 10^{11-12}$GeV, which has been obtained for the case of
$\alpha=0$. We have, however, found that similar values of the Hubble
constant $H$ are obtained even for $\alpha\ne 0$ as long as $\left|
  \alpha \right|\lesssim 1$. From Eq.(\ref{fa-alpha}) we derive $F_a
\theta_a \gtrsim 10^{15}$GeV for $\alpha \lesssim 0.1$ and $H\simeq
10^{11-12}$GeV. Reversely, if $\alpha$ takes a value of ${\cal
  O}(10^{-2})$ (for example, $\alpha\simeq 0.05$) as we have seen in
\S\ref{sec:obs}, and $H \simeq 10^{11}$GeV, we have $F_{a}\theta_a
\simeq 1.5\times 10^{15}$GeV.

In the standard cosmology, $F_a$ has the upper limit $F_a\lesssim
10^{12}$GeV.~\cite{Kolb-Turner} However, as shown in Ref. (28)
this constraint is greatly relaxed if late-time entropy production
takes place.  In this case, the unclosure condition for the present
universe leads to an upper bound for $F_a \theta_a$ as
\begin{equation}
  F_a \theta_a \lesssim 4.4\times 10^{15}{\rm GeV},
\end{equation}
where the reheating temperature after late-time entropy production is
taken as $T_R = 1$MeV. This relaxed constraint is consistent with
$F_a$ ($\simeq 10^{15}$~GeV) required for $\alpha \simeq 0.05$ and
$H \simeq 10^{11-12}$~GeV.

\subsection{New inflation model}

In this subsection, we consider the new inflation model proposed by
Izawa and Yanagida,~\cite{Izawa-Yanagida} which is based on an $R$
symmetry in supergravity.

In this model, the inflaton superfield $\phi(x, \theta)$ is assumed to
have an $R$ charge $2/(n+1)$ so that the following tree-level
superpotential is allowed:
\begin{equation}
        W = -\frac{g}{n+1}\phi^{n+1},
        \label{sup-pot}
\end{equation}
where $n$ is a positive integer and $g$ denotes a coupling constant of
order $1$. We further assume that the continuous $U(1)_R$ symmetry is
dynamically broken down to a discrete $Z_{2nR}$ at a scale $v$,
generating an effective superpotential:~\cite{Izawa-Yanagida,Kumekawa}
\begin{equation}
        W_{eff} = v^{2}\phi - \frac{g}{n+1}\phi^{n+1}.
        \label{sup-pot2}
\end{equation}

%
%

The $R$-invariant effective K\"ahler potential is given by
\begin{equation}
        K(\phi) = |\phi|^{2} + \frac{\zeta}{4} |\phi|^{4} + \cdots,
        \label{kahler}
\end{equation}
where $\zeta$ is a constant of order $1$. As shown in Ref. 29),
the spectral index $n_s$ of the density fluctuations is given by
\begin{equation}\label{new-index}
  n_s \simeq 1-2\zeta.
\end{equation}
By using the experimental constraint $n_s \gtrsim 0.8$, we take a
relatively small value for $\zeta$, $\zeta\lesssim 0.1$.

Let us now discuss the inflationary dynamics of the new inflation
model.  We identify the inflaton field $\varphi(x)/\sqrt{2} (\ge 0)$
with the real part of the scalar component field of superfield $\phi$.
(We use the same symbol for a scalar component field as the
superfield.) The potential for the inflaton is given by
\begin{equation}
  V(\varphi) \simeq v^4 - \frac{\zeta}{2}v^4\varphi^2 -
  \frac{g}{2^{\frac{n}{2}-1}}v^2\varphi^n +
  \frac{g^2}{2^n}\varphi^{2n}
\end{equation}
for $\varphi < \langle\varphi \rangle = \sqrt{2}\langle \phi \rangle =
\sqrt{2} \left( v^2/g \right)^{1/n}$. Here, $g$ and $v$ are taken to
be positive. It is shown in Ref. 29)
that the slow-roll condition for the inflaton is satisfied for $\zeta
< 1$ and $\varphi \lesssim \varphi_f$, where
\begin{equation}
  \varphi_f \simeq \sqrt{2}\left(\frac{(1-\zeta)v^2}{gn(n-1)}
  \right)^{\frac{1}{n-2}}.
\end{equation}
This provides the value of $\varphi$ at the end of inflation.
Hereafter we take $n=4$, since it is shown in Ref. 29)
that this is the most plausible case.

\begin{figure}
\centerline{\psfig{figure=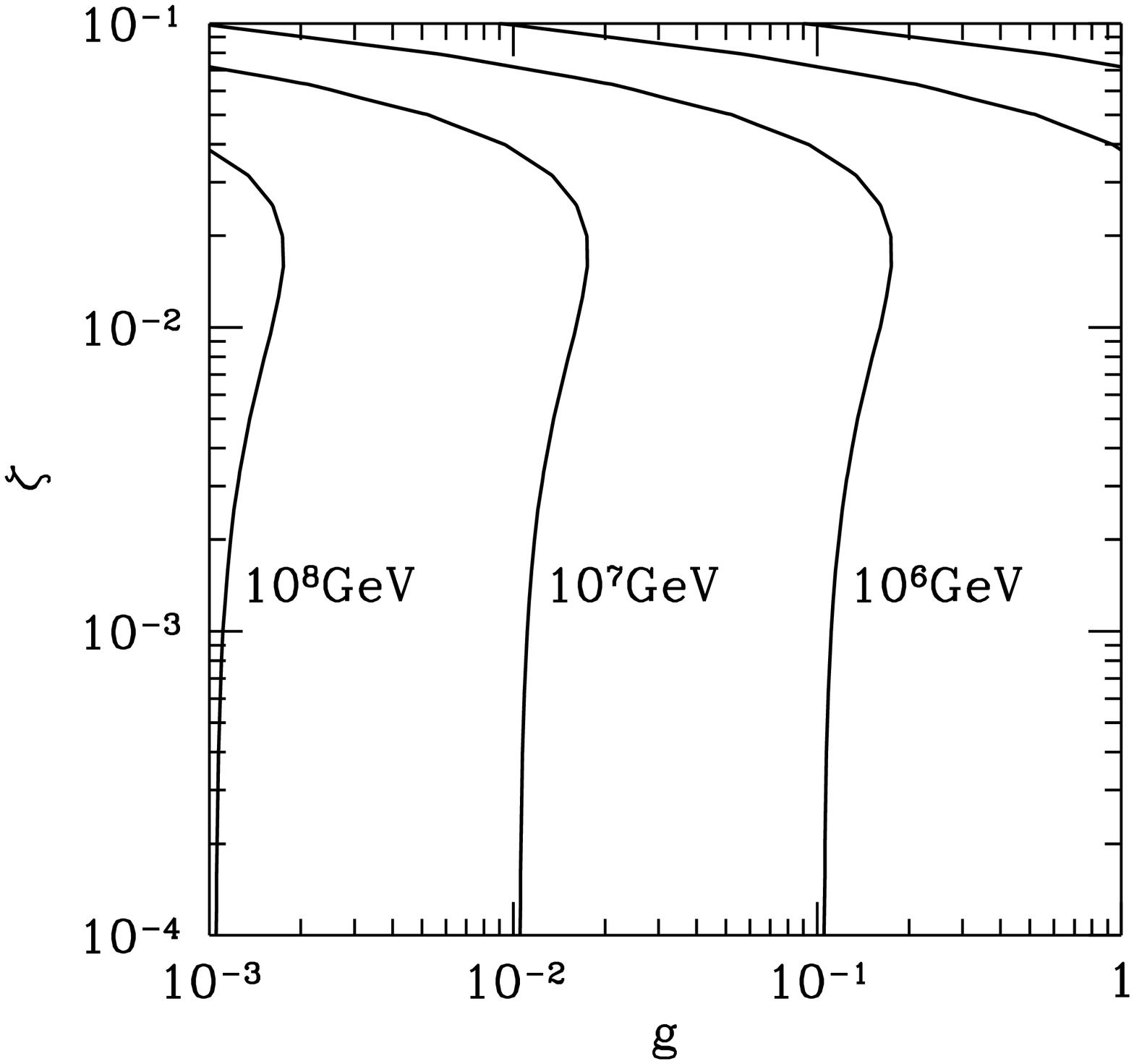,height=15cm} }
    \vspace{1cm}
    \caption{The Hubble parameter during the new inflation, which is
      normalized by the COBE, ignoring tensor perturbations and
      isocurvature fluctuations.}
  \label{fig:newhubble_inflation}
\end{figure}

We assume that the inflaton begins rolling down its potential from
near the origin ($\varphi\approx 0$). The Hubble parameter during the
inflation ($0 < \varphi \lesssim \varphi_f$) remains almost constant
and is given by $H = V^{1/2}/\sqrt{3} \simeq v^2/\sqrt{3}$. When
$\varphi$ reaches $\varphi_f$, the inflation ends.  As in the hybrid
case, we need an e-folding number $N\sim 60$, and we solve the
equation of motion numerically. Also, we can fix one parameter $v$ by
using Eq.(\ref{iso-cobe}). The result is that $H$ is as large as
$H\sim {\cal O}(10^{6-7})$GeV with the reasonable parameter region
$10^{-3} \lesssim g \lesssim 1$, $10^{-4} \lesssim \zeta \lesssim
10^{-1}$ (see Fig.\ref{fig:newhubble_inflation}). This value is for
$\alpha=0$, but similar values are obtained for $\alpha \lesssim 0.1$.
If we take $\alpha \sim {\cal O}(10^{-2})$ and $H\sim {\cal
  O}(10^{7-8})$GeV, we have from Eq.(\ref{fa-alpha}) $F_a \theta_a
\sim 10^{12}$GeV, which is the canonical value for axionic dark matter
without late-time entropy production.\footnote{For large $\zeta\ 
  (\gtrsim 0.05)$, the spectral index deviates from $1$ (see
  Eq.(\ref{new-index})), and the observational constraint on $\alpha$
  discussed in \S\ref{sec:obs} becomes slightly stringent. However,
  the result in \S\ref{sec:obs} can be directly applied for smaller
  $\zeta$.}

\subsection{Chaotic inflation model}

It is difficult to construct a realistic model of chaotic inflation in
supergravity. However, a monomial type of chaotic inflation model is
very simple and is treated widely in the literature. We consider such
a model in this subsection.

For simple models of chaotic inflation such as
\begin{equation}
V(\phi) = \frac{\lambda \phi^4}{4}
\end{equation}
or 
\begin{equation}
V(\phi) = \frac{m}{2}\phi^2,
\end{equation}
the COBE normalized value of Hubble constant during inflation is about
$H\sim {\cal O}(10^{14})$GeV. In this case, a too great amount of
isocurvature fluctuations is produced, even considering late-time
entropy production. However, if the potential for Peccei-Quinn scalar
field is extremely flat, the effective value of $F_a$ during inflation
can be larger and $F_a \sim 10^{18}$GeV is possible.~\cite{LindePQ}
When this is the case, we have an interesting amount of isocurvature
fluctuations again. For example, if $F_a\simeq 2.4\times 10^{18}$GeV
and $\alpha \simeq 0.05$, we have $H\sim 1.4\times 10^{14}$GeV, which
is a natural value in chaotic inflation.\cite{KSY}

\section{Conclusions and discussion}
\label{sec:conclusion}

In this paper we have studied density fluctuations which have both
adiabatic and isocurvature modes and have discussed their effects on
the large scale structure of the universe and CMB anisotropies. By
comparing the observations of the large scale structure, we have found
that the mixed fluctuation model is consistent with observations if
the ratio $\alpha$ of the power spectrum of isocurvature fluctuations
to that of adiabatic fluctuations is less than $\sim 0.1$. In
particular, the mixed fluctuation model with $\alpha \sim 0.05$, total
matter density $\Omega_{0} =0.4$, vacuum energy density $\lambda_{0} =
0.6$, and Hubble parameter $H_0=70$km/s/Mpc gives a very good fit to
both the observations of the abundance of clusters and the shape
parameter. (In recent observations of high-redshift supernovae, a
similarly large value for $\lambda_{0}$ has been
obtained.~\cite{Riess}) Therefore, the mixture model of isocurvature
and adiabatic fluctuations is astrophysically interesting.

The CMB anisotropies induced by the isocurvature fluctuations can be
distinguished from those produced by pure adiabatic fluctuations
because the shapes of the angular power spectrum of CMB anisotropies
are quite different from each other on small angular scales. The most
significant effect of the mixture of isocurvature fluctuations is that
the acoustic peak in the angular power spectrum decreases.  From
observations of the CMB anisotropies in future satellite experiments,
we may know to what degree the axionic isocurvature fluctuations are
present.

The requirement $\alpha \lesssim 0.1$ leads to constraints on the
Peccei-Quinn scale $F_a$ and the Hubble parameter during the
inflation. Isocurvature fluctuations with $\alpha \simeq 0.05$ are
produced naturally in the hybrid inflation model if we take the
Peccei-Quinn scale $F_a \simeq 10^{15-16}$~GeV. This $F_a$ exceeds the
usual upper bound ($F_a \lesssim 10^{12}$~GeV). However, this bound
can be relaxed if there exists late-time entropy production, and in
this case $F_a \sim 10^{15-16}$~GeV is allowed. Furthermore, $F_a
\simeq 10^{16}$~GeV is predicted in the M-theory.~\cite{Banks} Thus
the existence of large isocurvature fluctuations may provide crucial
support for the M-theory axion
hypothesis.~\cite{Kawasaki-Yanagida,KKSY}

In the new inflation model, we also have sufficiently large
isocurvature fluctuations to be observed. In this case, we do not have
to invoke late-time entropy production to dilute the axion density. In
a simple chaotic inflation model, we have a larger value for $H$
during inflation than the above two cases. In this case, $F_a$ exceeds
the upper limit, even considering the late-time entropy production.
However, the Peccei-Quinn scalar field may have an extremely flat
potential, and $F_a$ may be as large as the gravitational scale. In
this case, we have an appropriate amount of isocurvature fluctuations
again.

In all of these three realistic classes of inflation models, we have
astrophysically interesting amounts of isocurvature fluctuations.

\acknowledgements

One of the authors (T.K.) is grateful to K. Sato for his continuous
encouragement and to T. Kitayama for useful discussions.

\begin{thebibliography}{99}
\bibitem{KSY} M.  Kawasaki, N. Sugiyama, and T. Yanagida, 
    Phys. Rev. {\bf D54}, 2442 (1996).
\bibitem{Stomper} R.~Stomper, A.~J.~Banday and K.~M.~G\'orski,
  Astrophys. J. {\bf 463}, 8 (1996).
\bibitem{Burns} S.~D.~Burns, {\tt astro-ph/9711303}.
\bibitem{Scherrer} A.~A.~de Laix and R.~J.~Scherrer, Astrophys. J.
  {\bf 464}, 539 (1996).
\bibitem{Peacock} J. A. Peacock and S. J. Dodds, 
    Mon. Not. Roy. Astron. Soc. {\bf 267}, 1020 (1994).
  \bibitem{KKSY}T. Kanazawa, M. Kawasaki, N.~Sugiyama and T.~Yanagida,
    Prog. ~Theor. ~Phys. {\bf 100}, 1055 (1998).
\bibitem{BBKS} J.M. Bardeen, J.R. Bond, N. Kaiser and A.S. Szalay
    Astrophys. J. {\bf 304}, 15 (1986).
\bibitem{Sugiyama} N. Sugiyama, 
    Astrophys. J. Supp. {\bf 100}, 281 (1995).
\bibitem{Davis} M. Davis and P. J. E. Peebles, 
    Astrophys. J. {\bf 267}, 465 (1983).
\bibitem{COBE-norm} C.L.~Bennett et al.,  
    Astrophys. J. Letter {\bf 464}, L1 (1996);
    E. Bunn and M. White,
    Astrophys. J. {\bf 480}, 6 (1997).
\bibitem{KSBEHS} H. Kodama and M. Sasaki
    Int. J. Mod. Phys. A{\bf 1}, 265 (1986);
    G. Efstathiou and J.R. Bond,
    Mon. Not. R. Astron. Soc. {\bf 227}, 33 (1987);
    W. Hu and N. Sugiyama,
    Phys. Rev. {\bf D51}, 2599 (1995).
  \bibitem{Eke} V. R. Eke, S. Cole and C. S. Frenk, Mon. Not. Roy.
    Astron. Soc. {\bf 282}, 263 (1996).
\bibitem{White} S. D. M. White, G. Efstathiou and C. S. Frenk,
    Mon. Not. Roy. Astron. Soc. {\bf 262}, 1023 (1993).
\bibitem{Viana} P. T. P. Viana and A. R. Liddle, 
    Mon. Not. Roy. Astron. Soc. {\bf 281}, 323 (1996).
  \bibitem{Kitayama} T. Kitayama and Y. Suto, Astrophys. J. {\bf 490},
    557 (1997).
\bibitem{Freedman} 
    W.L. Freedman, {\tt astro-ph/9706072}.
  \bibitem{MAP} http://map.gsfc.nasa.gov/
  \bibitem{PLANCK}
    http://astro.estec.esa.nl/SA-general/Projects/Planck/
\bibitem{MFP} V.~F.~Mukhanov, H.~A.~Feldman and R.~H.~Brandenberger,
  Phys.~Rep.~{\bf 215}, 203 (1992).
\bibitem{Kodama-Sasaki} H. Kodama and M. Sasaki, Int. J. Mod. Phys.
  {\bf A1}, 265 (1986); {\bf A2}, 491 (1987).
  \bibitem{Linde-Riotto} A. Linde and A. Riotto, Phys. Rev. {\bf D56},
    1841 (1997).
  \bibitem{Wess-Bagger} J.~Wess and J.~Bagger, {\it Supersymmetry and
      Supergravity} (Princeton University Press, Princeton NJ, 1992).
  \bibitem{Copeland} E. J. Copeland et al., Phys. Rev. {\bf D49}, 6410
    (1994).
  \bibitem{Dvali} G. Dvali, Q. Shafi and R. Schaefer, Phys. Rev. Lett.
    {\bf 73}, 1886 (1994).
  \bibitem{Coleman-Weinberg} S. Coleman and E. Weinberg, Phys. Rev.
    {\bf D7}, 1888 (1973).
  \bibitem{Panagiotakopoulos} C.~Panagiotakopoulos, Phys.~Lett.~{\bf
      B402}, 257 (1997).
  \bibitem{Kolb-Turner} E. W. Kolb and M. S. Turner, {\it The Early
      Universe} (Addison-Wesley, Reading MA, 1990).
  \bibitem{KMY} M. Kawasaki, T. Moroi and T. Yanagida, Phys. Lett.
    {\bf B383}, 313 (1996).
  \bibitem{Izawa-Yanagida} Izawa K-. I. and T. Yanagida, Phys. Lett.
    {\bf B393}, 331 (1997).
  \bibitem{Kumekawa} K. Kumekawa, T. Moroi and T. Yanagida, Prog.
    Theor. Phys. {\bf 92}, 437 (1994).
  \bibitem{LindePQ} A. D. Linde, Phys. Lett. {\bf B259}, 38 (1991).
  \bibitem{Riess} A.~G.~Riess et al., {\tt astro-ph/9805201}.
\bibitem{Banks} T.~Banks and M.~Dine, 
    Nucl.~Phys.~{\bf B479}, 173 (1996); 
    Nucl.~Phys.~{\bf B505}, 445 (1997).
\bibitem{Kawasaki-Yanagida} M.~Kawasaki and T.~Yanagida,
    Prog.~Theor.~Phys.~{\bf 97}, 809 (1997).
\end{thebibliography}
\end{document}